\documentclass[paper]{JHEP}
\usepackage{epsfig}
\def\as{\alpha_s}
\def\a0{\bar\alpha_0}
\def\ae{\alpha_{\mbox{\scriptsize eff}}}
\def\aPT{\as^{\mbox{\tiny PT}}}
\def\ee{e^+e^-}
\def\eps{\epsilon}
\def\cA{{\cal A}}
\def\cF{{\cal F}}
\def\cG{{\cal G}}
\def\cM{{\cal M}}
\def\cV{{\cal V}}
\def\half{{\textstyle {1\over2}}}
\def\mI{\mu_{\mbox{\tiny I}}}
\def\mR{\mu_{\mbox{\tiny R}}}
\def\MSbar{{$\overline{\mbox{\rm MS}}$}}
\def\VEV#1{\langle{#1}\rangle}
\def\re#1{(\ref{#1})}
\def\beq{\begin{equation}}   \def\eeq{\end{equation}}
\def\beeq{\begin{eqnarray}}   \def\eeeq{\end{eqnarray}}
\title{\boldmath Two-loop enhancement factor for $1/Q$
corrections to event shapes in deep inelastic scattering}
\author{Mrinal Dasgupta and Bryan R.\ Webber\thanks{Research
supported in part by the U.K. Particle Physics and Astronomy Research
Council and by the EC Programme ``Training and Mobility of Researchers",
Network ``Hadronic Physics with High Energy Electromagnetic Probes",
contract ERB FMRX-CT96-0008.}\\
        Cavendish Laboratory, University of Cambridge,\\
        Madingley Road, Cambridge CB3 0HE, U.K.\\
        E-mail: \email{dasgupta@hep.phy.cam.ac.uk}, etc.}
\abstract{We compute the two-loop enhancement factors for
our earlier one-loop calculations of leading ($1/Q$) power
corrections to the mean values of some event shape variables
in deep inelastic lepton scattering.  The enhancement is
found to be equal to the universal ``Milan factor'' for
those shape variables considered, provided the one-loop
calculation is performed in a particular way. As a result,
the phenomenology of power corrections to DIS event shapes
remains largely unaffected.}
\preprint{Cavendish--HEP--98/02\\hep-ph/9809247}
\begin{document}
\section{Introduction}
In a recent paper \cite{DasWeb98} we presented a theoretical study
of the leading power-suppressed corrections to the mean values of
various event-shape measures in deep inelastic scattering (DIS).
Our investigation was based on an analysis of one-loop Feynman
graphs containing a massive gluon, which probes the sensitivity
of different observables to long-distance
physics \cite{Web94,BBNNA,Neu} and is closely related to the
analysis of infrared renormalons \cite{Muel,Zakh,KorSte,AkZak}.
We found that, as in $\ee$ annihilation, the leading corrections
to DIS event shapes are normally of order $1/Q$. This prediction
has since been verified experimentally \cite{H1}.

By making the further assumption of universal low-energy behaviour
of the strong coupling, we also estimated the magnitudes
of $1/Q$ corrections to event shapes in DIS relative
to those observed in $\ee$ annihilation \cite{DELPHI,JADEOPAL}.
Here the one-loop massive-gluon analysis suffers from some
ambiguities and deficiencies. The ambiguities arise from
the sensitivity of some event shapes to the way in which
the gluon mass is included in the definition of the shape
variable, the phase space and the matrix elements. This problem
is related to the inadequacy of the massive-gluon approach
pointed out by Nason and Seymour \cite{NaSey}: unlike the
total $\ee$ annihilation cross section, for example, event
shapes are sensitive to the precise way in which a virtual
gluon fragments into observable particles.  In other words,
they are not sufficiently inclusive to be expressible in terms
of the distribution of an effectively massive gluon.

In a two-loop analysis of this problem in $\ee$ event
shapes, however, the Milan group \cite{DLMS} has found that,
for a wide class of shape variables, the effects of
non-inclusiveness amount simply to an enhancement
of the ``naive massive-gluon'' estimate of the $1/Q$
correction by a universal ``Milan factor'', with
numerical value $\cM\simeq 1.49$. The two-loop analysis
also clarified the way in which the gluon mass should be
included at the ``naive'' level, in order that
universality should be manifest. 

In the present paper we perform the same type of two-loop
analysis for event shape variables in DIS. We find
a universal enhancement of the $1/Q$ corrections to
the current jet thrust, mass and $C$-parameter,
with the same enhancement factor $\cM$,
provided the ``naive massive-gluon'' estimate is
computed in a particular way. This resolves the
ambiguities we encountered in the one-loop calculation
of power corrections to certain DIS shape variables.

In the case of the current jet broadening variable, which
we also studied at one-loop level in Ref.~\cite{DasWeb98},
there are additional complications arising from recoil
contributions \cite{broad}, and therefore we postpone
the two-loop analysis to a future publication.

The layout of the present paper is as follows.
In Sect.~\ref{sec_kin}
we define convenient kinematic variables for the study of
multi-parton emission in DIS. These variables are used
in  Sect.~\ref{sec_shapes} to express the various event
shape variables that we investigate. We shall argue that
the dominant two-loop contributions to the leading
power correction come from the kinematic region in
which the emitted gluons are soft. The contributions
from this region are described in Sect.~\ref{sec_glsplit}.
Sect.~\ref{sec_BPY} explains the `dispersive' method
we use to calculate the power corrections. The actual
calculations are then performed in Sect.~\ref{sec_pow},
and finally the results are summarized and discussed in
Sect.~\ref{sec_conc}.

\section{Kinematics}\label{sec_kin}

\EPSFIGURE{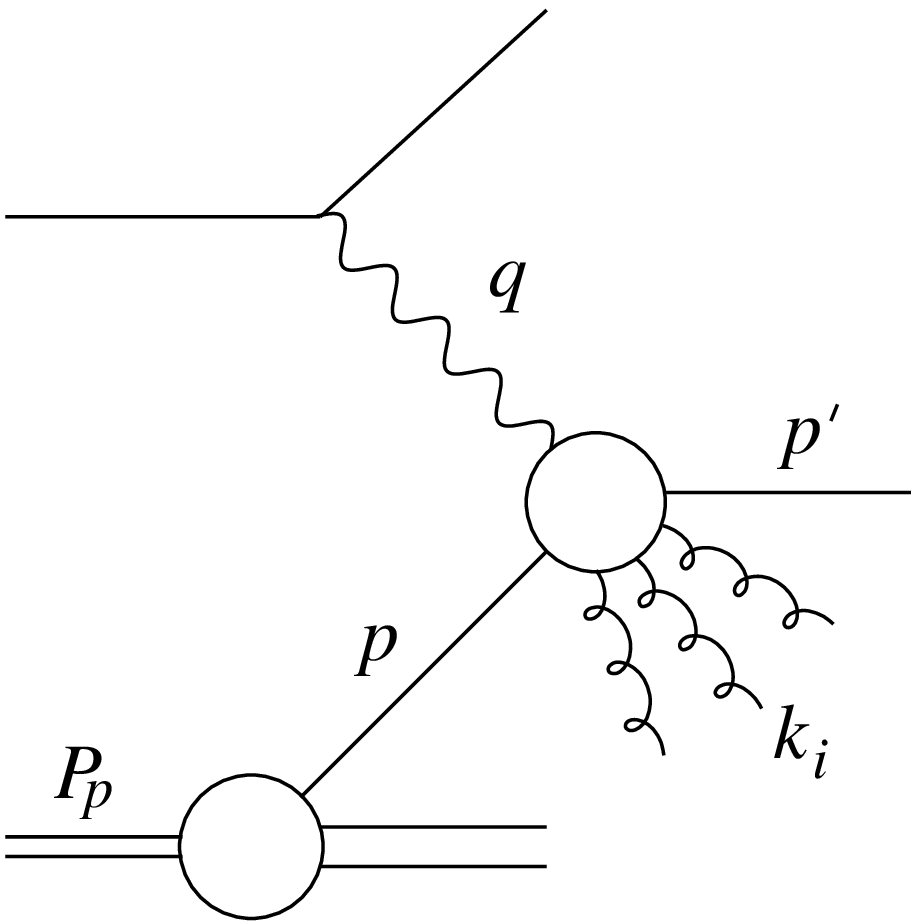}{Deep inelastic scattering with
multi-parton emission.}

It is convenient to write the momenta $k_i$ of radiated partons
(gluons and/or quark-antiquark pairs) in terms of Sudakov
(light-cone) variables as (Fig.~1)
\beq
 k_i\>=\>  \alpha_i P \>+\>\beta_i P' \>+\>k_{ti}\>,
\qquad\qquad \alpha_i\beta_i\>=\>\vec k_{ti}^2/Q^2
\eeq
where $P$ and $P'$ are light-like vectors
along the incoming parton and current directions,
respectively, in the Breit frame of reference:
\beq
P\>=\>x P_p\>,\>\>
P'\>=\>x P_p+q\>,\>\> 
2(P\cdot P')\>=\>-q^2\>\equiv\>Q^2
\eeq
where $P_p$ is the incoming proton momentum and
$x=Q^2/2(P_p\cdot q)$ is the Bjorken variable. 
Thus in the Breit frame we can write
$P=\half Q(1,0,0,-1)$ and $P'=\half Q(1,0,0,1)$,
taking the current direction as the $z$-axis. Particles
in the {\em current hemisphere} $H_c$ have $\beta_i>\alpha_i$
while those in the proton {\em remnant hemisphere} $H_r$ have
$\alpha_i>\beta_i$.

The momenta $p$ and $p'$ of the initial- and final-state quarks
can also be resolved along the Sudakov vectors  $P$ and $P'$.
From momentum conservation,
\beeq
p  &=& (1+\alpha'+\sum\alpha_i)P\nonumber \\
p' &=& \alpha' P+(1-\sum\beta_i)P'-\sum k_{ti}
\eeeq
where
\beq\label{alprime}
\alpha'(1-\sum\beta_i)\>=\>\left(\sum\vec k_{ti}\right)^2/Q^2\;.
\eeq
The initial quark is assumed to be collinear with the proton
direction and is therefore aligned along $P$. It may have an
``intrinsic'' transverse momentum of the order of the hadronic
mass scale $\lambda\sim 1$ GeV, but this will lead to
corrections of order $\lambda^2/Q^2$, which we neglect.
The final quark momentum $p'$ is mostly along $P'$; its
component $\alpha'$ along $P$ is small. In perturbation
theory, $\alpha'$ is of order $\as(Q^2)$ on
the average, while in the soft region (all $k_{ti}\sim\lambda$)
it is of order $\lambda^2/Q^2$. When some emissions are hard and
some soft, the non-perturbative contributions to event shapes
are at most of order $\as(Q^2)\lambda/Q$,
which we also neglect. Thus, as far as $\lambda/Q$ corrections
are concerned, $\alpha'$ can be regarded as contributing only
to the perturbative part of event shapes, and we can
compute the non-perturbative part from a study of the
soft region alone.

In the soft region, the matrix elements for multi-parton
emission in DIS will be the same as those for $\ee$ annihilation
into quark-antiquark plus soft partons at c.m.\ energy-squared
$Q^2$, with the outgoing antiquark replaced by an incoming quark
of momentum $p$. This means that we shall be able to take the
results for various matrix element integrals directly from the
study of $\ee$ event shapes \cite{DLMS}, adjusted according to
the definition of DIS event shapes.

\section{DIS event shape variables}\label{sec_shapes}
\subsection{Current jet thrust}
Consider first the current jet thrust $T_Q$, defined by the
sum of longitudinal momenta in the current hemisphere
normalized to $Q/2$:
\beq
T_Q =\frac 2 Q \sum_{a\in H_c}p_{za}\>=\>
\frac 2 Q \left(p'_z+\sum_{H_c} k_{zi}\right)\;.
\eeq
We have assumed that the outgoing quark momentum $p'$ lies in
the current hemisphere, because the probability that this
is not true is of order $\as(Q^2)$, and the associated
non-perturbative contribution will be at most of order
$\as(Q^2)\lambda/Q$, which we neglect. In the notation
of the previous section, we have
\beeq\label{tauQ}
\tau_Q\equiv 1-T_Q
&=& \sum\beta_i -\sum_{H_c}(\beta_i-\alpha_i)+\alpha'
\nonumber \\
&=& \sum_{H_r}\beta_i +\sum_{H_c}\alpha_i+\alpha'
\>=\>\sum\min\{\alpha_i,\beta_i\}+\alpha'\;.
\eeeq
Where not explicitly stated, the sums extend over all partons,
irrespective of which hemisphere they are in. Here the contribution
of those partons that are not directly in the current hemisphere
represents the loss of momentum of the outgoing quark.

Note that the expression \re{tauQ} is {\em additive}, in
the following sense: the mean value of the event shape
receives a set of perturbative contributions, from $\alpha'$
and $\min\{\alpha_i,\beta_i\}$ for hard parton emissions, plus
a non-perturbative contribution from soft emissions, which
can be computed from the expression
\beq\label{tauQsoft}
\tau_Q
\>\simeq\>\sum\min\{\alpha_i,\beta_i\}\;,
\eeq
valid in the soft region, and added to the perturbative part.
  
If instead we measure the thrust with respect to
an axis chosen to maximize its value in the current hemisphere,
we obtain the same expression in the soft region, since the
transverse contribution is of second order in the small
quantities $k_{ti}/Q$.

Another possible way of defining the current jet thrust is to
normalize it to the total energy in the current hemisphere,
instead of $Q/2$. This gives
\beq
T_E = T_Q/(1-\varepsilon)
\eeq
where 
\beq\label{epsdef}
\varepsilon  \equiv 1-\frac 2 Q \sum_{a\in H_c}E_a\;.
\eeq
The quantity $\varepsilon$, which we call the {\em energy deficit}
in the current hemisphere, is itself an interesting shape variable,
not measured up to now.  It is given by
\beq\label{epsbet}
\varepsilon
= \sum\beta_i -\sum_{H_c}(\beta_i+\alpha_i)-\alpha'
\>=\>\sum_{H_r}\beta_i -\sum_{H_c}\alpha_i-\alpha'\;.
\eeq
and hence in the soft region
\beq\label{TEsof}
\tau_E\equiv 1-T_E\simeq 2\sum_{H_c}\alpha_i \;.
\eeq

\subsection{Current jet mass}
The current jet mass is defined by
\beq\label{rhodef}
\rho=\left(\sum_{a\in H_c}p_a\right)^2/Q^2\;.
\eeq
In the notation of Sect.~\ref{sec_kin},
\beq\label{rho}
\rho=\left(\alpha'+\sum_{H_c}\alpha_i\right)\cdot
\left(1-\sum_{H_r}\beta_i\right)-\sum_{H_r}\alpha_i\beta_i\;.
\eeq
Once again we should ignore $\alpha'$ and terms of second
order in $\alpha_i$ and/or $\beta_i$ when computing $1/Q$
corrections, in which case we have simply the additive soft
contribution
\beq\label{rhosof}
\rho\>\simeq\>\sum_{H_c}\alpha_i\;.
\eeq

\subsection{$C$-parameter}
The $C$-parameter is
\beq\label{Cdef}
C=6\sum_{a,b\in H_c}|\vec p_a||\vec p_b|\sin^2\theta_{ab}/Q^2\;.
\eeq
Note the factor of 6 replacing the 3/2 that appears in the
definition of the $C$-parameter for $\ee$ annihilation,
because of the normalization to $(Q/2)^2$ in place of $Q^2$.

In the notation of Sect.~\ref{sec_kin}, the general expression
for the $C$-parameter is complicated,
but in the soft region we obtain simply
\beq\label{Csof}
C\>\simeq\>12\sum_{H_c}\frac{\alpha_i\beta_i}{\alpha_i+\beta_i}\;,
\eeq
which is again additive.

\section{Soft two-loop contributions}\label{sec_glsplit}
Consider now the contribution to a shape variable V when
two soft, not necessarily collinear partons are emitted.
Following \cite{DLMS}, we write the two-parton phase space as
\beq
d\Gamma_2(k_1,k_2) = 
\frac{d\alpha_1}{\alpha_1}\frac{d\alpha_2}{\alpha_2}
\frac{d^2k_{t1}}{\pi}\frac{d^2k_{t2}}{\pi} = 
\frac{d\alpha}{\alpha}\frac{d^2k_t}{\pi}\frac{d\phi}{2\pi}dz\,dm^2
\eeq
where
\beeq
  \alpha&=&\alpha_1+\alpha_2\>, \quad 
   \alpha_1=z\alpha\>, \>\> \alpha_2=(1-z)\alpha\>,\nonumber \\
  m^2 &=& (k_1+k_2)^2\>, \quad \vec k_t = \vec k_{t1}+\vec k_{t2}
\eeeq
and $\phi$ is an azimuthal angle.
The contribution to the mean value of V from one or two soft
partons, which will be denoted by $\cV$, may be written in the form
\beeq\label{cRV}
\cV &=&
{4C_F} \int\frac{d\alpha}{\alpha} \frac{d^2k_{t}}{\pi k_{t}^2}
\>\left(\frac{\as(0)}{4\pi} +\chi(k_{t}^2)\right)\> v_1(k)  \nonumber\\
&+&2C_F \int
d\Gamma_2(k_1,k_2)\left(\frac{\as}{4\pi}\right)^2
(C_FM_F+C_AM_A+n_fM_f)v_2(k_1,k_2)\,,
\eeeq
where $v_1(k)$ is the contribution to the shape variable from a
(massless) parton of momentum $k$ and $v_2(k_1,k_2)$ is the
corresponding two-parton contribution.

The first term on the right-hand side of Eq.~\re{cRV} represents
the lowest-order contribution of a single soft gluon (involving the
ill-defined quantity $\as(0)$) plus the two-loop virtual corrections
to it. The second term represents the contribution from the emission
of two soft partons, $M_F$ etc.\ being the parts of the
relevant matrix elements-squared with the corresponding
colour/flavour factors.  Both terms are divergent, but we
can rewrite $\cV$ as sum of collinear and infrared finite parts
by the following procedure. We introduce a term representing
a contribution from the combined parton momenta, $v_1(k_1+k_2)$, and
split the  two-parton contribution into three parts:
\beq
  v_2(k_1,k_2)\>=\> v_1(k_1+k_2)
\>+\> \left[\, v_1(k_1)+v_1(k_2)-v_1(k_1+k_2)\,\right]
\>+\> \left[\, v_2(k_1,k_2)-v_1(k_1)-v_1(k_2)\,\right]\;.  
\end{equation}
The first term defines the ``naive massive-gluon''
contribution, which treats the parton emission inclusively. 
The first expression in square brackets is the ``non-inclusive''
correction \cite{DLMS}. The second square bracket defines
what we shall call the ``non-additive'' contribution.

The combined momentum $k_1+k_2$ is massive and therefore
the definition of the contribution $v_1(k_1+k_2)$ is somewhat
ambiguous. It does not matter how we define this quantity, as long
as the expression used becomes equal to $v_1(k_1)+v_1(k_2)$ in the soft
and/or collinear (i.e.\ massless) limit. We define it as follows: if
\beq
v_1(k_i)\>=\> v(\alpha_i,\beta_i)\>, \qquad
\alpha_i\beta_i=k_{ti}^2/Q^2\>,
\eeq
then
\beq
v_1(k_1+k_2)\>\equiv\> v(\alpha,\beta)\>, \qquad
\alpha=\alpha_1+\alpha_2\>,\>\>
\beta=\beta_1+\beta_2\>; \quad \alpha\beta=(k_t^2+m^2)/Q^2\>.
\eeq

The expression \re{cRV} may now be written as the sum of four
finite parts, as follows \cite{DLMS}:
\beq\label{cVsplit}
 \cV \>=\> \cV_0 + \cV_{in} + \cV_{ni} + \cV_{na}\>.
\eeq
The first term is the so-called ``naive'' contribution.
It incorporates only those two-loop terms proportional to the
beta-function coefficient $\beta_0=(11C_A-2n_f)/3$,
which together with $\as(0)$ build up the running coupling,
as will be explained in the following section:
\beq\label{cV0}
\cV_0 \> \equiv\>
4C_F \int \frac{dm^2 dk^2_t}{k^2_t+m^2}
\left\{\frac{\as(0)}{4\pi}\delta(m^2)
-\frac{\beta_0}{m^2}\left(\frac{\as}{4\pi}\right)^2\right\}
\Omega_0\left((k_t^2+m^2)/Q^2\right)
\eeq
where $\Omega_0$, the ``naive trigger function'', is given by
\beq\label{0trigger}
  \Omega_0(\eps)\>\equiv\> \int_\eps^1 \frac{d\alpha}{\alpha} 
  \,v\left(\alpha,\,\beta=\eps/\alpha\right)\;.
\eeq

The second term in Eq.~\re{cVsplit} is the ``inclusive'' correction,
which combines the remaining non-Abelian part of the virtual correction
with the corresponding ``naive massive gluon'' part of the two-parton
contribution. This was shown in Ref.~\cite{DLMS} to be given by
\beeq\label{cVin}
\cV_{in} &=&
8C_F C_A\int \frac{dm^2 dk^2_t}{m^2(k^2_t+m^2)}
\left(\frac{\as}{4\pi}\right)^2
\ln\left[\frac{k_t^2(k_t^2+m^2)}{m^4}\right]\nonumber \\
& &\cdot\left[\Omega_0\left((k_t^2+m^2)/Q^2\right)
-\Omega_0(k_t^2/Q^2)\right]
\eeeq

The ``non-inclusive'' correction is the additive part of the
remaining non-Abelian (and quark-antiquark) two-parton contribution:
\beq\label{cVni}
\cV_{ni} \> =\>
\frac{C_F}{\pi}\int dm^2\,dk_t^2\,dz\,d\phi
\left(\frac{\as}{4\pi}\right)^2 (C_AM_A+n_fM_f)\,
\Omega_{ni}(m^2/Q^2,k_t^2/Q^2,z,\phi)
\eeq
where the ``non-inclusive trigger function'' is
\beq\label{nitrigger}
\Omega_{ni}(m^2/Q^2,k_t^2/Q^2,z,\phi)\>\equiv\>
\int_{(k_t^2+m^2)/Q^2}^1\frac{d\alpha}{\alpha}
\;[\, v_1(k_1)+v_1(k_2)-v_1(k_1+k_2)\,]\,.
\eeq
Note that the Abelian ($C_F^2$) part of this contribution
is cancelled by virtual corrections \cite{DLMS}.

Finally the ``non-additive'' correction is the rest of the
two-parton contribution:
\beq\label{cVna}
\cV_{na} \> =\>
\frac{C_F}{\pi}\int dm^2\,dk_t^2\,dz\,d\phi
\left(\frac{\as}{4\pi}\right)^2 (C_FM_F+C_AM_A+n_fM_f)\,
\Omega_{na}(m^2/Q^2,k_t^2/Q^2,z,\phi)
\eeq
where the ``non-additive trigger function'' is
\beq\label{natrigger}
\Omega_{na}(m^2/Q^2,k_t^2/Q^2,z,\phi)\>\equiv\>
\int_{(k_t^2+m^2)/Q^2}^1\frac{d\alpha}{\alpha}
\;[\, v_2(k_1,k_2)-v_1(k_1)-v_1(k_2)\,]\,.
\eeq
All the quantities considered in Sect.~\ref{sec_shapes} are
additive, in the sense that $v_2(k_1,k_2)=v_1(k_1)+v_1(k_2)$
in the soft region, and so there is no contribution $\cV_{na}$
in these cases. 

\section{Dispersive calculation of power corrections}\label{sec_BPY}
Our method for estimating power corrections will
assume that the QCD coupling $\as(k^2)$ can be defined down to
arbitrarily low values of the scale $k^2$ and that it has reasonable
analytic properties, i.e.\ no singularities other than a cut
along the negative real axis. It follows that one can write the
formal dispersion relation
\beq\label{alphas}
\as(k^2) = -\int_0^\infty \frac{dm^2}{m^2+k^2}\;\rho_s(m^2)
\eeq
where the spectral function $\rho_s$ represents the discontinuity
across the cut,
\beq\label{rhos}
\rho_s(m^2) = \frac{1}{2\pi i}\mbox{Disc}\left\{\as(-m^2)\right\}
\equiv \frac{1}{2\pi i}\left\{\as\left(m^2 e^{i\pi}\right)
-\as\left(m^2 e^{-i\pi}\right)\right\}\;.
\eeq

Our strategy will be to rewrite the results of the previous
section in terms of $\rho_s(m^2)$ and then to
study the effect of a non-perturbative modification to
this function in the soft region.  We eliminate $\as(0)$ from
Eq.~\re{cV0} using Eq.~\re{alphas}, and insert $\rho_s(m^2)$
in the second-order terms in place of $\as^2$ using the result
\beq
\rho_s = -\frac{\beta_0}{4\pi}\as^2 - \ldots\;.
\eeq
The appropriate scale for the argument of $\rho_s$ in this
substitution is $m^2$, since the two-loop contribution
in Eq.~\re{cV0} generates the running coupling. We assume
that the same argument can be used in Eqs.~\re{cVin} and
\re{cVni}. Then, for example, Eq.~\re{cV0} becomes
\beq\label{cV0rho}
\cV_0 \>=\> \int\frac{dm^2}{m^2}\,\rho_s(m^2)\,\cF_0(m^2/Q^2)
\eeq
where the naive {\it characteristic function} $\cF_0(m^2/Q^2)$ is
\beq\label{cF0}
\cF_0(m^2/Q^2) = \frac{C_F}{\pi}\int_0^{Q^2}
dk^2_t\left[\frac{\Omega_0((k_t^2+m^2)/Q^2)}{k^2_t+m^2}
-\frac{\Omega_0(k_t^2/Q^2)}{k^2_t}\right]\;.
\eeq
Similarly we may define  $\cF_{in}$ etc.\ for the inclusive,
non-inclusive and non-additive corrections,
\beeq\label{cFin}
\cF_{in}(m^2/Q^2) &=&
-\frac{2C_F C_A}{\pi\beta_0}\int \frac{dk^2_t}{k^2_t+m^2}
\ln\left[\frac{k_t^2(k_t^2+m^2)}{m^4}\right]
\left\{\Omega_0\left((k_t^2+m^2)/Q^2\right)
-\Omega_0(k_t^2/Q^2)\right\}\nonumber \\ \label{cFni}
\cF_{ni}(m^2/Q^2) &=&
-\frac{C_Fm^2}{4\pi^2\beta_0}\int dk_t^2\,dz\,d\phi\,
(C_AM_A+n_fM_f)\,\Omega_{ni}(m^2/Q^2,k_t^2/Q^2,z,\phi)\;
\\ \label{cFna}
\cF_{na}(m^2/Q^2) &=&
-\frac{C_Fm^2}{4\pi^2\beta_0}\int dk_t^2\,dz\,d\phi\,
(C_FM_F+C_AM_A+n_fM_f)\,\Omega_{na}(m^2/Q^2,k_t^2/Q^2,z,\phi)
]\nonumber\;,
\eeeq
so that
\beq\label{cVrho}
\cV \>=\> \int\frac{dm^2}{m^2}\,\rho_s(m^2)\,\cF(m^2/Q^2)
\eeq
where $\cF$ is the two-loop corrected characteristic function,
\beq
\cF = \cF_0+\cF_{in}+\cF_{ni}+\cF_{na}\;.
\eeq

Non-perturbative effects at long distances are expected to give rise
to a modification in the strong coupling at low scales, $\delta\as$,
which generates a corresponding modification in the
spectral function via Eq.~\re{rhos}:
\beq\label{deltas}
\delta\rho_s(m^2)
= \frac{1}{2\pi i}\mbox{Disc}\left\{\delta\as(-m^2)\right\}\;.
\eeq

Inserting this in Eq.~\re{cVrho} and rotating the integration
contour separately in the two terms of the discontinuity, we obtain
the following non-perturbative contribution to Eq.~\re{cVrho}:
\beq\label{deltaF}
\delta\cV = \int_0^\infty \frac{dm^2}{m^2}\,\delta\as(m^2)
\,\cG(m^2/Q^2)
\eeq
where, setting $m^2/Q^2 = \eps$,
\beq\label{Gdef}
\cG(\eps) = -\frac{1}{2\pi i}\mbox{Disc}\left\{\cF(-\eps)\right\}\;.
\eeq
Since $\delta\as(m^2)$ is limited to low values of
$m^2$, the asymptotic behaviour of $\delta\cV$ at large
$Q^2$ is controlled by the behaviour of $\cF(\eps)$ as
$\eps\to 0$.  We see from Eq.~\re{Gdef} that no terms analytic
at $\eps=0$ can contribute to $\delta\cV$.  On the
other hand for a square-root behaviour at small $\eps$,
\beq\label{deltaF1}
\cF\sim a_V \frac{C_F}{2\pi}\sqrt{\eps}
\qquad\Longrightarrow\qquad
\delta\cV = -\frac{a_V}{\pi}\frac{\cA_1}{Q}\;,
\eeq
where $\cA_1$ is the $q=1$ moment of the strong coupling
modification \cite{DLMS,DasSmyWeb},
\beq\label{cAdef}
\cA_q \equiv \frac{C_F}{2\pi}
\int_0^\infty \frac{dm^2}{m^2}\,m^q\,\delta\as(m^2)\;.
\eeq
Note that $\delta\as$ is a different quantity from the
``effective coupling modification'' $\delta\ae$ introduced
in Ref.~\cite{BPY}. Their moments are related by the
formula \cite{DasSmyWeb}
\beq\label{AcArel}
A_q \equiv \frac{C_F}{2\pi}
\int_0^\infty \frac{dm^2}{m^2}\,m^q\,\delta\ae(m^2)
\>=\> \frac{\sin(\pi q/2)}{\pi q/2}\cA_q\;.
\eeq
Thus in particular $A_1 = 2 \cA_1/\pi$,
and all the even moments of $\delta\ae$ are zero.

We can express $\cA_1$ in terms of the average value of $\as$
in the infrared region, as follows \cite{DokWeb95,DokWeb97}. 
We substitute for $\delta\as$ in Eq.~\re{cAdef}
\beq
\delta\as(m^2) \simeq \as(m^2) - \aPT(m^2)\;,
\eeq
where $\aPT$ represents the expression for $\as$
corresponding to the part already included in the
perturbative prediction.  As discussed in Ref.~\cite{DokWeb95},
if the perturbative calculation is carried out to second
order in the \MSbar\ renormalization scheme, with renormalization
scale $\mR^2$, then we have
\beq
\aPT(m^2) = \as(\mR^2) + [b\ln(\mR^2/m^2)+k]\,\as^2(\mR^2) 
\eeq
where for $N_f$ active flavours ($C_A=3$)
\beq
b = \frac{11 C_A-2N_f}{12\pi}\;,\;\;\;\;
k= \frac{(67-3\pi^2)C_A-10N_f}{36\pi}\;.
\eeq
The constant $k$ comes from a change of scheme from \MSbar\ to
the more physical scheme \cite{CMW} in which $\as$ is preferably
defined at low scales.
Then above some infrared matching scale $\mI$ we assume that
$\as(m^2)$ and $\aPT(m^2)$ approximately coincide, so that
\beeq\label{A1exp}
\cA_1 &\simeq& \frac{C_F}{\pi}
\int_0^{\mI} dm\,\left(\as(m^2)-
\as(\mR^2) - [b\ln(\mR^2/m^2)+k]\,\as^2(\mR^2)\right)\nonumber\\
&=& \frac{C_F}{\pi}\,\mI\,\left(\a0(\mI)-
\as(\mR^2) - [b\ln(\mR^2/\mI^2)+k+2b]\,\as^2(\mR^2)\right)\;,
\eeeq
where
\beq\label{a0def}
\a0(\mI) \equiv \frac{1}{\mI}\int_0^{\mI}\as(m^2)\,dm\;.
\eeq
Studies of event shapes in $\ee$ annihilation \cite{DELPHI,JADEOPAL}
suggest that $\a0\simeq 0.5$ for $\mI=2$ GeV, which translates into
a value of $\cA_1\simeq 0.2$ GeV for $Q\sim 20-100$ GeV.

\section{Power corrections to DIS event shapes}\label{sec_pow}
We now use the method of the previous section to evaluate the
coefficients of $1/Q$ for the DIS shape variables defined in
Sect.~\ref{sec_shapes}.
\subsection{Current jet mass}
Consider first the current jet mass, which is given in the soft
region by the simplest expression, from Eq.~\re{rhosof}:
\beq\label{vrho}
v(\alpha,\beta) = \alpha\,\Theta(\beta-\alpha)\;.
\eeq
Eq.~\re{0trigger} therefore gives for the naive trigger function
\beq\label{omrho}
 \Omega_0^{(\rho)}(\eps)\>=\>\sqrt{\eps}\;,
\eeq
dropping all terms of order $\eps$ (i.e.\ order $1/Q^2$).

In writing Eq.~\re{vrho} we neglected the fact that the
momentum fraction of the struck quark is not $x$ but $x(1+\alpha)$.
If we normalize to the Born cross section at fixed $x$, this means
there should be an extra factor of $q(x(1+\alpha))/q(x)$ on the
right-hand side, where $q(x)$ is the relevant quark distribution.
However, by Taylor expansion we see that this factor only gives
a correction of order $\eps$, and so we may ignore it.

Substituting Eq.~\re{omrho} in Eq.~\re{cF0}, we obtain
\beq\label{cF0rho}
\cF_0^{(\rho)}(\eps) = -2\frac{C_F}{\pi}\,\sqrt{\eps}\;.
\eeq
By simple dimensional analysis, the inclusive and non-inclusive
corrections to the characteristic function, given by Eqs.~\re{cFin}
and \re{cFni}, are also proportional to $\sqrt\eps$:
\beq\label{cFinrho}
\cF_{in}^{(\rho)}(\eps) = -2\frac{C_F}{\pi}\,r_{in}\sqrt{\eps}
\eeq
where
\beq\label{rin}
r_{in} =\frac{C_A}{\beta_0}\int_0^\infty
\frac{2x\ln[x^2(1+x^2)]}{(1+x^2)(x+\sqrt{1+x^2})}dx
\>=\> 3.2994\frac{C_A}{\beta_0}\;,
\eeq
and
\beq\label{cFnirho}
\cF_{ni}^{(\rho)}(\eps) = -2\frac{C_F}{\pi}\,r_{ni}\sqrt{\eps}
\eeq
where
\beq\label{rni}
r_{ni} = \frac{m}{8\pi\beta_0}\int dk_t^2\,dz\,d\phi\,
(C_AM_A+n_fM_f)\,(k_{t1}+k_{t2}-\sqrt{k_t^2+m^2})
\eeq
with
\beeq
k_{t1}^2 &=& z[zk_t^2+(1-z)m^2+2\sqrt{z(1-z)}\,k_tm\,\cos\phi]\;,
\nonumber \\
k_{t2}^2 &=& (1-z)[(1-z)k_t^2+zm^2-2\sqrt{z(1-z)}\,k_tm\,\cos\phi]\;.
\eeeq
The integration in Eq.~\re{rni} was performed for $\ee$ annihilation
in Ref.~\cite{DLMS}. As explained above, the result for DIS will
be identical in the soft region, and therefore we can
take the value directly from there (which we also confirmed
numerically):
\beq\label{rnival}
r_{ni} = -\frac{2}{\beta_0}(0.862C_A + 0.052 n_f)\;.
\eeq
Since the contributions to the jet mass from multiple soft parton emission
are simply additive, see Eq.~\re{rho}, there is no non-additive
correction. The two-loop corrected characteristic function
for the current jet mass is thus
\beq\label{cFrho}
\cF^{(\rho)}(\eps) = -2\cM\frac{C_F}{\pi}\,\sqrt{\eps}\
\eeq
where $\cM$ is the {\em Milan factor} computed in Ref.~\cite{DLMS},
\beeq\label{Milan}
\cM &=& 1+r_{in}+r_{ni} = 1+(1.575C_A - 0.104n_f)/\beta_0\nonumber \\
 &=& 1.49\>\>\mbox{for}\;n_f=3\;.
\eeeq

It follows from Eqs.~\re{deltaF1} and \re{cFrho} that the leading
power correction to the mean current jet mass is
\beq\label{drhonew}
\delta\VEV\rho = \frac{4\cM}{\pi}\frac{\cA_1}{Q}
\>=\>1.89\frac{\cA_1}{Q}\>\>\mbox{for}\;n_f=3\;.
\eeq
Comparing with our earlier result given in Ref.~\cite{DasWeb98},
\beq\label{drhoold}
\delta\VEV\rho \sim 2\frac{A_1}{Q}\;,
\eeq
we note the following differences:
\begin{itemize}
\item The first moment of the effective coupling modification,
$A_1$, is replaced by $2\cA_1/\pi$ where $\cA_1$ is the first
moment of the modification to the strong coupling itself,
in accordance with Eq.~\re{AcArel}.
\item The result is enhanced by a factor which is identical to
the Milan factor obtained in $\ee$ annihilation \cite{DLMS}.
\end{itemize}
This means that universality of $1/Q$ corrections still holds,
in the sense that the coefficient of $1/Q$ for the DIS current jet
mass remains one half of that for the thrust in $\ee$ annihilation,
for example. Furthermore the numerical coefficient of $\cA_1/Q$
is only slightly lower than the
coefficient of $A_1/Q$ calculated earlier using the ``naive
massive gluon'' approximation, because the Milan factor $\cM$ is
largely cancelled by the translation factor of $2/\pi$.
Thus experimental results \cite{H1} interpreted earlier as
measurements of $A_1$ using the ``naive'' formula can be
reinterpreted as giving a similar value for $\cA_1$,
taking into account the Milan and translation factors.

\subsection{Current jet thrust}
For the quantity $\tau=1-T_Q$ where $T_Q$ is the current jet
thrust, we have from Eq.~\re{tauQsoft}
\beq\label{vT}
v(\alpha,\beta) = \alpha\,\Theta(\beta-\alpha)
+\beta\,\Theta(\alpha-\beta)\;,
\eeq
and Eqs.~\re{0trigger} and \re{cF0} now give
\beq
 \Omega_0^{(\tau)}(\eps)\>=\>2\sqrt{\eps}
\;,\>\>\cF_0^{(\tau)}(\eps)\>=\>-4\frac{C_F}{\pi}\,\sqrt{\eps}\;.
\eeq
Similarly, both the inclusive and non-inclusive corrections to the
characteristic function are twice as large as those for the jet
mass, and again there is no non-additive part,
so the leading power correction is
\beq\label{taunew}
\delta\VEV\tau = \frac{8\cM}{\pi}\frac{\cA_1}{Q}
\>=\>3.79\frac{\cA_1}{Q}\>\>\mbox{for}\;n_f=3\;,
\eeq
to be compared with our earlier result of $4A_1/Q$ in Ref.~\cite{DasWeb98}.
Once again the two-loop result \re{taunew} has a similar
coefficient, but $A_1$ is replaced by$\cA_1$.

As stated earlier, the above result for the $1/Q$
power correction (unlike the perturbative contribution)
does not depend on whether the thrust is measured relative to
the current axis or an axis which maximizes its value.
We also see from Eq.~\re{TEsof} that the same leading power
correction should be obtained if the thrust is normalized to the
total energy in the current hemisphere, instead of $Q/2$.  This
is because the mean energy deficit in the current hemisphere,
$\varepsilon$, defined in Eq.~\re{epsdef}, should have no $1/Q$ correction,
since the two terms in Eq.~\re{epsbet} cancel on the
average.\footnote{We expect the leading non-perturbative correction
to $\VEV{\varepsilon}$ to be of order $\as(Q^2)/Q$.}

\subsection{$C$-parameter}
Similarly for the $C$-parameter \re{Cdef}, we have
\beq\label{vC}
v(\alpha,\beta) = 12\frac{\alpha\beta}{\alpha+\beta}\Theta(\beta-\alpha)
\eeq
and the naive trigger function is
\beq
 \Omega_0^{(C)}(\eps)\>=\>3\pi\sqrt{\eps}\;,
\eeq
giving
\beq\label{cF0C}
\cF_0^{(C)}(\eps) = -6C_F\,\sqrt{\eps}\;.
\eeq
The inclusive and non-inclusive corrections scale in the same way
and again there is no non-additive part, so
\beq\label{Cnew}
\delta\VEV C = 12\cM\,\frac{\cA_1}{Q}
\>=\>17.88\frac{\cA_1}{Q}\>\>\mbox{for}\;n_f=3\;,
\eeq
Our one-loop result for $C$ in Ref.~\cite{DasWeb98} was ambiguous,
since it depended on how the gluon mass was included. Following the
definition used here, the ``naive'' result is $6\pi A_1/Q = 18.8 A_1/Q$.
So again the two-loop result \re{Cnew} has a similar coefficient,
but with $A_1$ replaced by $\cA_1$.

\section{Conclusions}\label{sec_conc}
\TABLE{
\begin{tabular}{|c|c c|}\hline
$\cV$        & 1-loop & 2-loop \\ \hline
$\VEV{\rho}$ &    2   &  1.9 \\  
$\VEV{\tau}$ &    4   &  3.8 \\  
$\VEV{C}$    &  $6\pi-12\pi$   &  17.9 \\ \hline  
\end{tabular}
\caption{Coefficients of $A_1/Q$ (1-loop) and  $\cA_1/Q$ (2-loop).}
}
Our results on the coefficients of the leading power corrections to
the DIS event shapes defined in Sect.~\ref{sec_shapes} are
summarized in Table 1. For all the shape variables studied,
we have found that, as in $\ee$ annihilation \cite{DLMS}, the
enhancement factor for the leading power correction is equal
to the Milan factor \re{Milan}, provided the ``naive''
coefficient is computed in the way specified in
Sect.~\ref{sec_glsplit}. This eliminates the ambiguity
in the prediction for the $C$-parameter,
which was present at the 1-loop level \cite{DasWeb98} as
indicated.

In all cases the enhancement factor practically cancels the
suppression factor of $2/\pi$ that comes from translating the
first moment of the effective coupling modification, $A_1$,
into the corresponding quantity for the strong coupling itself,
$\cA_1$. This quantity can be used to infer the value of $\a0$,
the low-energy moment of $\as$ defined in Eq.~\re{a0def}.

The experimental data on thrust and jet masses in both
DIS \cite{H1}\footnote{The thrust and jet mass values we
refer to here are called $T_c$ and $\rho_c$ in Ref.~\cite{H1}.
For these quantities the coefficients used there to fit the data
were close to those given in Table 1.}
and $\ee$ annihilation \cite{DELPHI,JADEOPAL}
suggest a value of $\a0\simeq 0.5$. Final experimental
results on the $C$-parameter in DIS have yet to be presented.

\acknowledgments
We are most grateful to Yu.L.\ Dokshitzer, A.\ Lucenti, G.\ Marchesini
and G.P.\ Salam, and also to members of the H1 Collaboration, especially
H.-U.\ Martyn and K.\ Rabbertz, for many helpful discussions and comments.
M.D.\ acknowledges the financial support of Trinity College, Cambridge 
and the comments and contributions 
of G.E. \ Smye and L. \ Magnea, the co-authors of 
Ref.~\cite{DasMagSmy}, where an independent calculation led to the 
discovery and rectification of an error in the numerical value 
of $\cM$.

\end{document}